\documentstyle[11pt]{article}

\begin{document}

\begin{center}
{\large \bf Asymptotic behavior and critical coupling in the 
scalar Yukawa model from Schwinger-Dyson equations}\\
\vspace*{5 true mm}
{\bf V.E. Rochev\footnote{E-mail address: rochev@ihep.ru}}\\
{\it Institute for High Energy Physics, 142280 Protvino, Russia}
\end{center}
{\small{ {\bf Abstract.} A sequence of
$n$-particle approximations for the system of Schwinger-Dyson
equations is
investigated in the
model of a complex scalar field $\phi$
and a real scalar field $\chi$ with the interaction
$g\phi^*\phi\chi$.
 In the first non-trivial two-particle approximation, the system 
is reduced to
a system of two nonlinear integral equations for  propagators.
 The study of this system
shows that for equal masses a
critical coupling constant $g^2_c$ exists,
which separates the weak- and strong-coupling regions with 
 the different asymptotic behavior for  deep Euclidean momenta.
In the weak-coupling region ($g^2<g^2_c$), the propagators are
asymptotically  free, which corresponds to the wide-spread opinion
about the dominance of  perturbation theory for this model.
At the critical point the asymptotics  of  propagators are $\sim 1/p$.
 In the 
strong coupling region  ($g^2>g^2_c$), the propagators are
asymptotically constant, which corresponds to the
ultra-local limit.  For unequal masses,  the critical point transforms into 
a segment of values, in which there are no solutions with a self-consistent
ultraviolet behavior without Landau singularities.
}\\


\newpage

\section{Introdution}

A system of  Schwinger-Dyson equations (SDEs) is the infinite set of integral
equations containing, in principle, all the information about
 a model of quantum field theory.
So far, there are no effective methods for the study of this
infinite system as a whole, so it is necessary to truncate
this system with a finite set of equations. This truncation
usually refers to either an expansion in a small parameter
(examples of such expansions are the coupling-constant perturbation theory
 and the $1/N$-expansion), or a simulation (and
rather, guessing) of some properties of the model (see, e.g., \cite{Swanson}).
An initial truncation of the system determines 
further approximations of the complete system, which is thus
attached to the leading approximation.

The system of SDEs is, in fact, the system of relations between derivatives
of the generating functional of Green functions, resulting from the
  functional-differential
SDE, which acts as a dynamical principle of the theory.
If we approximate the generating functional with the first $n$ terms
of 
an expansion in powers of a source,
the system of SDEs
can be approximated by a closed system of  integral equations. This system defines
the $n$-th term of a sequence of approximations, which
for $n\rightarrow \infty$ obviously goes into
the complete system of SDEs. A sequence of such approximations
 in the model of  complex scalar field with
interaction $\lambda(\phi^*\phi)^2$ was considered
in \cite{Rochev11}.
For the simplest nontrivial
approximation (the ``two-particle approximation'')
an  asymptotic solution of the corresponding system  in deep Euclidean
region  was obtained.  In the strong-coupling region, this solution 
is  free from  Landau singularities. 
 In \cite{Rochev12},
in the same approximation, the Yukawa model was considered, for which
 a self-consistent solution without 
Landau  singularities
 in the Euclidean region also has been constructed.

In this paper, in the same approximation, we consider a system
of SDEs in the  model of the complex scalar field $\phi$ (phion)
  and real scalar field $\chi$ (chion)
with the interaction $g\phi^*\phi\chi$ in four dimensions.
  This model, also known
 as a scalar Yukawa model, is used in nuclear
physics as a simplified version of the Yukawa model without spin
degrees of freedom, as well as an effective model of
the interaction of scalar quarks \cite{Guasch}.
 Despite its
 well-known imperfection  associated with
its instability \cite{Baym} (or more precisely, the metastability
\cite{Cornwall} -- \cite{Savkli}), this model, as
the simplest model of the interaction of fields, often used
as a prototype of more substantive theories to elaborate
the  various non-perturbative approaches in
the quantum field theory.

In the  two-particle approximation, the  system of SDEs
for the  scalar Yukawa model is
a system of two nonlinear integral equations for  propagators.
 The study of this system
shows 
a change of the asymptotic behavior of  propagators
in the deep Euclidean region in a vicinity of
 a certain critical value of the coupling constant.
For small values of the coupling the
propagators behave as free, which is consistent with the
wide-spread opinion  about the dominance of  perturbation theory
for this super-renormalizable model. In the strong-coupling 
region, however, the asymptotic behavior changes 
dramatically -- both propagators in the deep Euclidean region
tend to some constant limits.

The existence  of a critical coupling constant in the
scalar Yukawa model was noticed by practically all authors
who
have investigated this model using different methods (see, e.g.,
\cite{Nieuwen}-\cite{Sauli} and references therein).
This critical constant is generally regarded as
a limit on the coupling constant
for a self-consistent
description of the model by some method.
In our approach, however, the self-consistent solution
for propagators exists also for the strong coupling,
and the existence of the critical coupling
  looks more like as a phase
transition in accordance
with the general definition of the phase transition as a sharp
change of properties of the model with a smooth change of parameters
(see, e.g., \cite{Glimm}).

The structure of the paper is as follows. In section 2, the general
 formalism of SDEs for the scalar Yukawa model is described
and the necessary 
definitions and notations are given.  A
simplest non-perturbative expansion, namely the mean-field
expansion is also considered
in this section. The leading term of this expansion corresponds to
the chain summation for a chion propagator.
The problem of restoring of crossing properties 
for this expansion is discussed, 
and the existence of a critical
value $g^2_c$  of the coupling  is shown.
For $g^2>g^2_c$, the
asymptotics  of inverse chion  propagator in deep Euclidean region becomes negative,
which leads to the appearance of a  Landau singularity 
 and the associated violation of the self-consistency of the method.

In section 3,  a construction of 
$n$-particle approximations for the system of SDEs is given,
and the two-particle approximation is considered in detail.
 The system of SDEs in the two-particle approximation is reduced
to  a system of nonlinear integral equations for propagators.
The study of the system at large
Euclidean momenta shows that for the equal masses of fields
 a self-consistent solution exists for each
value of the coupling, and there is a critical coupling value,
 which separates the weak and strong coupling regions
with a different asymptotic behavior. For the unequal masses,
the weak and strong coupling regions are separated by the 
segment of the intermediate coupling without
 self-consistent solutions.

In section 4, 
 a three-particle approximation for
the model is briefly  considered, and 
 the crossing-symmetry
problem for
  two-particle amplitude is discussed.
 A discussion of the results is contained in
section 5. In the Appendix, the exact solution of
linearized integral equation in the strong-coupling region
is obtained  for the case of
equal masses.

\section{SDEs and
the mean-field expansion 
in the bilocal source formalism}

\subsection{Preliminaries}

We consider the model of interaction of a
complex scalar field $\phi$ (phion) and a real scalar field $\chi$
(chion) with the Lagrangian
\begin{equation}
  {\cal L}= -\partial_\mu\phi^*\partial_\mu\phi-m^2_0\phi^*\phi-
\frac{1}{2}(\partial_\mu\chi)^2-\frac{\mu^2}{2}\chi^2+g\phi^*\phi\chi
\end{equation}
 in a four-dimensional Euclidean space 
$(x\in E_4)$.  The coupling constant $g$ has a dimension of mass .

The generating functional of  Green functions (vacuum
averages) is the functional integral
\begin{equation}
G(\eta, j)=\int
D(\phi,\phi^*,\chi)\exp\bigg\{\int dx\, {\cal L}(x)-
\int dxdy\,\phi^*(y)\eta(y, x)\phi(x)+\int dx \,j(x)\chi(x)
\bigg\}.
\end{equation}
Here, $\eta$ is a bilocal source of the phion field,\footnote{The formalism
  of the bilocal source 
 was first
elaborated in the quantum field theory 
by Dahmen and Jona-Lasinio \cite{Dahmen}.
 We consider this using  as a convenient choice
of the functional variable.}
 $j$ is a single source of the chion field.

 The translational invariance of the functional 
integration measure leads to  the functional-differential
SDEs for  the generating functional $G$. In terms of the  logarithm
 $Z=\log G$ these equations are 
\begin{equation}
g\bigg[\frac{\delta^2 Z}{\delta\eta(y,x)\delta
j(x)}+\frac{\delta Z}{\delta j(x)}\frac{\delta Z}{\delta \eta(y,x)}\bigg]
=(m^2_0-\partial_x^2)\frac{\delta Z}{\delta\eta(y,x)}+\int dx_1\,
\eta(x,x_1)\, \frac{\delta
Z}{\delta\eta(y,x_1)}+\delta(x-y),
\label{eta_eq}
\end{equation}
\begin{equation}
\frac{\delta Z}{\delta j(x)}=\int dx_1\,
D_c(x-x_1)\, j(x_1)-g\int dx_1\,D_c(x-x_1)\,
\frac{\delta Z}{\delta \eta(x_1,x_1)}.
\label{jeq}
\end{equation}
Here
$
D_c\equiv(\mu^2-\partial^2)^{-1}.
$

Equation  (\ref{jeq}) allows us to express
all  Green functions with chion legs in terms of functions that contain phions
only. Thus, the differentiation of (\ref{jeq}) over $\eta$ gives us
the three-point function
\begin{equation}
V(x,y|z)\equiv -\frac{\delta^2Z}{\delta j(z)\delta\eta(y,x)}\bigg|_{\eta=j=0}=
g\int dz_1\,
D_c(z-z_1)\, Z_2\left(\begin{array}{cc}z_1&z_1\\x&y\end{array}\right),
\label{3point}
\end{equation}
where
\begin{equation}
 Z_2\left(\begin{array}{cc}x&y\\x'&y'\end{array}\right)
\equiv\frac{\delta^2 Z}{\delta\eta(y',x')\delta\eta(y, x)}\,\bigg|_{\eta=j=0}
\end{equation}
is the two-particle phion function. 
The differentiation of  (\ref{jeq}) over  $j$ with
taking into account equation (\ref{3point}) gives us the chion
propagator:
\begin{equation}
 D(x-y)\equiv\frac{\delta^2 Z}{\delta j(y)\delta
j(x)}\bigg\vert_{\eta=j=0}
=D_c(x-y)+g^2\int dx_1dy_1\,
D_c(x-x_1)Z_2\left(\begin{array}{cc}x_1& x_1\\y_1&y_1\end{array}\right)
D_c(y_1-y),
\label{propchi}
\end{equation}
etc.

Excluding with  the help of the SDE (\ref{jeq}),
 a differentiation over $j$ in SDE (\ref{eta_eq}),
we obtain at $j=0$ the 
SDE for the generating functional:
\begin{eqnarray}
g^2\int dx_1\,
D_c(x-x_1)\,\bigg[\frac{\delta^2 Z}{\delta\eta(x_1,x_1)\delta\eta(y,x)}
+
\frac{\delta Z}{\delta \eta(x_1,x_1)}\,\frac{\delta Z}{\delta \eta(y,x)}\bigg]
+
\nonumber \\
+(m^2_0-\partial_x^2)\frac{\delta Z}{\delta\eta(y,x)}+
\int dy_1\,\eta(x,y_1)\,
\frac{\delta Z}{\delta\eta(y,y_1)}+\delta(x-y)=0,
\label{etaeq}
\end{eqnarray}
which only  contains
the derivatives over the bilocal
source $\eta$.

Sequential  differentiations of this equation give us
the infinite system of SDEs for Green functions.
For our purposes, it will need the first three equations of this system. 
Switching off the source
in equation (\ref{etaeq}), we obtain the equation
\begin{equation}
 (m^2-\partial^2_x)\Delta(x-y)=\delta(x-y)+
g^2\int dx_1\,
D_c(x-x_1)\,
Z_2\left(\begin{array}{cc}x&y\\x_1&x_1\end{array}\right).
\label{D1}
\end{equation}
Here
$
m^2\equiv m^2_0-\frac{g^2}{\mu^2}\Delta(x=0)
$
and
\begin{equation}
 \Delta(x-y)\equiv-\frac{\delta
Z}{\delta\eta(y,x)}\bigg\vert_{\eta=0}
\label{prophi}
\end{equation}
is the phion propagator. 
 
A differentiation over $\eta$ gives us the second equation:
\begin{eqnarray}
 (m^2-\partial^2_x)Z_2\left(\begin{array}{cc}x&y\\x'&y'\end{array}\right)-
g^2\int dx_1\,D_c(x-x_1)\, 
Z_2\left(\begin{array}{cc}x_1&x_1\\x'&y'\end{array}\right)\,\Delta(x-y)+
\nonumber \\
+g^2\int dx_1\,D_c(x-x_1)\, Z_3\left(\begin{array}{ccc}x_1&x_1\\x&y\\x'&y'\end{array}\right)
=\delta(x-y')\Delta(x'-y).
\label{D2}
\end{eqnarray}

The repeated  differentiation over $\eta$ gives one more equation:
\begin{eqnarray}
 (m^2-\partial^2_x)Z_3\left(\begin{array}{ccc}x&y\\x'&y'\\x''&y''\end{array}\right)-
g^2\int dx_1\,D_c(x-x_1)\,
 Z_3\left(\begin{array}{ccc}x_1&x_1\\x'&y'\\x''&y''\end{array}\right)
\,\Delta(x-y)+
\nonumber \\
+g^2\int dx_1\,D_c(x-x_1)\,
 Z_4\left(\begin{array}{cccc}x_1&x_1\\x&y\\x'&y'\\x''&y''\end{array}\right)
=-\bigg\{\delta(x-y'')Z_2\left(\begin{array}{cc}x'&y'\\x''&y\end{array}\right)+
\nonumber \\
+g^2\int dx_1\,D_c(x-x_1)\, Z_2\left(\begin{array}{cc}x_1&x_1\\x'&y'\end{array}\right)
\,Z_2\left(\begin{array}{cc}x''&y''\\x&y\end{array}\right)\bigg\}-
\bigg\{x'\leftrightarrow x'', y'\leftrightarrow y''\bigg\}.
\label{D3}
\end{eqnarray}
Here
$
Z_n\equiv\frac{\delta^n Z}{\delta\eta^n}\bigg|_{\eta=0}
$ is the $n$-particle phion function.

\subsection{Kernel and inverse kernel}

 Integral equations with the
kernel
\begin{equation}
 K_{ab}\left(\begin{array}{cc}x&y\\x'&y'\end{array}\right)=
\delta(x-x')\delta(y-y')-g^2\delta(x'-y')\int dx_1
\Delta_a(x-x_1)\,
D_c(x_1-y')\Delta_b(x_1-y)
\label{kernel}
\end{equation}
 will be repeatedly considered below  . 
Here $\Delta_a$ and $\Delta_b$ are the given functions.
It is easy to verify that the inverse  kernel has the form
\begin{equation}
 K_{ab}^{-1}\left(\begin{array}{cc}x&y\\x'&y'\end{array}\right)=
\delta(x-x')\delta(y-y')+\delta(x'-y')\int dx_1\Delta_a(x-x_1)\,
f(x_1-y')\Delta_b(x_1-y),
\label{inverse_kernel}
\end{equation}
i.e.
\begin{equation}
\int dx_1dy_1K_{ab}^{-1}\left(\begin{array}{cc}x&y\\x_1&y_1\end{array}\right)
\, K_{ab}\left(\begin{array}{cc}x_1&y_1\\x'&y'\end{array}\right)
=\delta(x-x')\delta(y-y') 
\end{equation}
(no summation over $a$ and $b$!).\\
In equation (\ref{inverse_kernel}), 
\begin{equation}
 f^{-1}(x-y)=\frac{1}{g^2}\,D_c^{-1}(x-y)-L_{ab}(x-y),
\label{feq}
\end{equation}
where
\begin{equation}
 L_{ab}(x-y)=\Delta_a(x-y)\Delta_b(y-x).
\label{L_ab}
\end{equation}
Note also  the useful  property of the
inverse kernel:
\begin{equation}
 g^2\int dx_1\,
D_c(x-x_1)\, K^{-1}_{ab}\left(\begin{array}{cc}x_1&x_1\\x'&y'\end{array}\right)
=f(x-y')\delta(x'-y').
\label{inv_k_prop}
\end{equation}

\subsection{Mean-field expansion}

The mean-field expansion for the generating functional
(see
\cite{Rochev11}, \cite{Rochev12} and  references therein)
\begin{equation}
 Z=Z^{(0)}+Z^{(1)}+\cdots
\end{equation}
is based on the leading approximation: 
\begin{equation}
 g^2 \int dx_1
D_c(x-x_1)\, \frac{\delta Z^{(0)}}{\delta\eta(x_1,x_1)}\,
\frac{\delta Z^{(0)}}{\delta\eta(y,x)}+
(m^2-\partial_x^2)\frac{\delta Z^{(0)}}{\delta\eta(y,x)}
 +\int dy_1 \eta(x,y_1)\, \frac{\delta
Z^{(0)}}{\delta\eta(y,y_1)}+\delta(x-y)=0.
\label{Z0}
\end{equation}
The NLO equation is
\begin{eqnarray}
g^2\int dx_1\,D_c(x-x_1)\, \frac{\delta Z^{(1)}}{\delta\eta(x_1,x_1)}\,
\frac{\delta Z^{(0)}}{\delta\eta(y,x)}+
g^2\int dx_1\,D_c(x-x_1)\, \frac{\delta Z^{(0)}}{\delta\eta(x_1,x_1)}\,
\frac{\delta Z^{(1)}}{\delta\eta(y,x)}+
\nonumber \\
+(m^2-\partial_x^2)\frac{\delta Z^{(1)}}{\delta\eta(y,x)}
 +\int dy_1\,\eta(x,y_1)\, \frac{\delta
Z^{(1)}}{\delta\eta(y,y_1)}
=-g^2\int dx_1\,D_c(x-x_1) \frac{\delta^2 Z^{(0)}}{\delta\eta(x_1,x_1)\delta\eta(y,x)},
\label{Z1}
\end{eqnarray}
etc.

Equation  (\ref{Z0}), after switching off the source, gives
 us the leading-order phion propagator. In the momentum space
\begin{equation}
 \Delta_{0}^{-1}(p)=m^2+p^2.
\label{Delta0}
\end{equation}

A differentiation of the leading-order equation over $\eta$ gives
us the equation for the leading-order two-particle function $Z_2^{(0)}$,
which can be written as:
\begin{equation}
\int dx_1dy_1 K_{00}\left(\begin{array}{cc}x&y\\x_1&y_1\end{array}\right)\,
Z_2^{(0)}\left(\begin{array}{cc}x_1&y_1\\x'&y'\end{array}\right)=
\Delta_0(x-y')\Delta_0(x'-y),
\end{equation}
where $K_{00}$ is kernel (\ref{kernel}) at $\Delta_a=\Delta_b=\Delta_0$.
In correspondence with equation (\ref{inverse_kernel}),
the solution of this equation is 
\begin{eqnarray}
 Z_2^{(0)}\left(
\begin{array}{cc} x&y\\x'&y'\end{array}
\right) 
=\Delta_0(x-y')\Delta_0(x'-y)+
\nonumber \\
+\int dx_1dx_2
\Delta_0(x-x_1)\Delta_0(x'-x_2)f_0(x_1-x_2)\Delta_0(x_1-y)\Delta_0(x_2-y'),
\label{Z20}
\end{eqnarray}
where
$
 f_0^{-1}=g^{-2}D_c^{-1}- L_0,
$
and
$
 L_0=\Delta_0\Delta_0
$
is the scalar loop. Equation (\ref{propchi}) after simple calculations
with taking into account above formulae 
gives us the leading-order chion propagator
   $D_0$. In the momentum space
\begin{equation}
 D_0^{-1}(p)=\mu^2+p^2-g^2L_0(p^2).
\label{D0}
\end{equation}

The repeated  differentiation of the leading-order equation over $\eta$
gives us the equation for the leading-order three-particle function
whose solution is
\begin{equation}
 Z_3^{(0)}\left(\begin{array}{ccc}x&y\\x'&y'\\x''&y''\end{array}\right)=
\int dx_1dy_1\,
K_{00}^{-1}\left(\begin{array}{cc}x&y\\x_1&y_1\end{array}\right)\,
Z_{30}\left(\begin{array}{ccc}x_1&y_1\\x'&y'\\x''&y''\end{array}\right).
\end{equation}
Here
\begin{eqnarray}
Z_{30}\left(\begin{array}{ccc}x&y\\x'&y'\\x''&y''\end{array}\right)\equiv
-\bigg\{\delta(x-y')Z_2^{(0)}\left(\begin{array}{cc}x'&y\\x''&y''\end{array}\right)+ 
\nonumber \\
+g^2\int dx_1
D_c(x-x_1)\, Z_2^{(0)}\left(\begin{array}{cc}x_1&x_1\\x'&y'\end{array}\right)
\,Z_2^{(0)}\left(\begin{array}{cc}x&y\\x''&y''\end{array}\right)\bigg\}-
\bigg\{x'\leftrightarrow x'', y'\leftrightarrow y''\bigg\},
\end{eqnarray}

Note the following property of the three-particle function resulting from
equation (\ref{inv_k_prop}):
\begin{equation}
 g^2\int dx_1\,D_c(x-x_1)Z_3^{(0)}\left(\begin{array}{ccc}x_1&x_1\\x&y\\x'&y'\end{array}\right)=
\int dx_1\,f_0(x-x_1)\,
Z_{30}\left(\begin{array}{ccc}x_1&x_1\\x&y\\x'&y'\end{array}\right).
\label{g2Z30}
\end{equation}

The other  functions of the leading approximation can be calculated
in  the same manner.

Note that in  contrast to the functional  derivatives of $Z$ over  single source
$j$, the higher derivatives of $Z$ over bilocal source $\eta$ 
are not connected parts of corresponding many-particle functions.
Thus, the two-particle phion function $Z_2$ is not
the connected part $Z_2^c$ of  two-particle  and
 related to it by the formula:
\begin{equation}
 Z_{2}\left(\begin{array}{cc}x&y\\x'&y'\end{array}\right)=
\Delta(x-y')\Delta(x'-y)+
Z_{2}^{c}\left(\begin{array}{cc}x&y\\x'&y'\end{array}\right)
\end{equation}

A characteristic feature of many-particle functions
 of the leading approximation is
their incomplete structure in terms of crossing symmetry.
 The Bose symmetry of the 
theory dictates the crossing symmetry of the connected part $ Z_2^c$:
\begin{equation}
 Z_{2}^c\left(\begin{array}{cc}x&y\\x'&y'\end{array}\right)=
Z_{2}^c\left(\begin{array}{cc}x'&y'\\x&y\end{array}\right)=
Z_{2}^c\left(\begin{array}{cc}x'&y\\x&y'\end{array}\right)=
Z_{2}^c\left(\begin{array}{cc}x&y'\\x'&y\end{array}\right).
\label{cross}
\end{equation}
It is easy  to see that $ Z_2^{(0)c}$ (this is the second term 
on the rhs of equation (\ref{Z20})) satisfies the first equality in
(\ref{cross}), but breaks the other two. This apparent discrepancy
is a feature of many non-perturbative approximations.
It is inherent, for example, to the Bethe-Salpeter equation in the ladder
approximation. We face  similar problems in the 
two-particle approximation, which will be considered below.
To restore the missing  crossing symmetry  of the
leading approximation, it is necessary to look at the 
next order.

Calculations in the next order quite similar to the above.
Equation (\ref{Z1}) with the source being switched off  gives us the equation
for the first correction
$\Delta_1$ to the propagator. The differentiation of
equation (\ref{Z1}) over the source gives the equation for the NLO two-particle
function $ Z_2^{(1)}$, which can be written as an
equation with kernel $K_{00}$, and taking into account the above
formulae the solution is 
\begin{equation}
 Z_2^{(1)}\left(\begin{array}{cc}x&y\\x'&y'\end{array}\right)
=Z_{21}\left(\begin{array}{cc}x&y\\x'&y'\end{array}\right)+
\int dx_1dx_2
\Delta_0(x-x_1)f_0(x_1-x_2)\Delta_0(x_1-y)
Z_{21}\left(\begin{array}{cc}x_2&x_2\\x'&y'\end{array}\right),
\end{equation}
where
\begin{eqnarray}
Z_{21}\left(\begin{array}{cc}x&y\\x'&y'\end{array}\right)\equiv 
-g^2\int dx_1dy_1
\Delta_0(x-y_1)D_c(y_1-x_1)
Z_3^{(0)}\left(\begin{array}{ccc}x_1&x_1\\y_1&y\\x'&y'\end{array}\right)+
\nonumber \\
+
\Delta_0(x-y')\Delta_1(x'-y)-
\frac{g^2}{\mu^2}\Delta_1(x=0)\int dx_1\Delta_0(x-x_1)
Z_2^{(0)}\left(\begin{array}{cc}x_1&y\\x'&y'\end{array}\right)+
\nonumber \\
+g^2\int dx_1dx_2
\Delta_0(x-x_1)D_c(x_1-x_2)\Delta_1(x_1-y)
Z_2^{(0)}\left(\begin{array}{cc}x_2&x_2\\x'&y'\end{array}\right).
\label{Z21}
\end{eqnarray}
From these equations, together with equation (\ref{g2Z30}), it follows that
$ Z_2^{(1)}$ contains the term
\begin{eqnarray}
\int dx_1dx_2
 \Delta_0(x-x_1)\Delta_0(x'-x_2)f_0(x_1-x_2)\Delta_0(x_2-y)\Delta_0(x_1-y')=
\nonumber \\
=Z_{2}^{(0)c}\left(\begin{array}{cc}x&y'\\x'&y\end{array}\right)=
Z_{2}^{(0)c}\left(\begin{array}{cc}x'&y\\x&y'\end{array}\right),
\end{eqnarray}
which restores the missing crossing symmetry of the LO two-particle
function.
Such restoration of the crossing symmetry is typical for
non-perturbative expansions in the formalism of bilocal
source (for similar examples in other models see
\cite{Rochev00}, \cite{Rochev09}).

The above equations contain divergent integrals
and require a renormalization.
  According to the standard recipe
we introduce the renormalized propagators $\Delta_r$ and $D_r$ and impose on them
the normalization condition\footnote{For the easement
 of the following calculations
 we choose the normalization point at zero
momenta.
Such a choice is not a big deal for  simple calculations of
the
mean-field expansion, but it is highly significant for
more complicated calculations in a more pithy
two-particle approximation (see below).}
\begin{equation}
 \Delta_r^{-1}(0)=m^2_r, \;\;
\frac{d\Delta_r^{-1}}{dp^2}\bigg|_{p=0}=1
\label{normDelta}
\end{equation}
and 
\begin{equation}
 D_r^{-1}(0)=\mu^2_r, \;\;
\frac{d D_r^{-1}}{dp^2}\bigg|_{p=0}=1.
\label{normD}
\end{equation}
In the leading approximation of the mean-field expansion
the phion propagator is given by equation
(\ref{Delta0}), which implies that it is sufficient to
replace
$
 \Delta_0\rightarrow\Delta_r, \; m^2_r\rightarrow m^2,
$
i.e., the renormalized leading-order phion propagator is
\begin{equation}
 \Delta_{r}^{-1}(p)=m_r^2+p^2.
\label{Delta0r}
\end{equation}
For the renormalized chion propagator 
according with equation (\ref{D0}),
we obtain in the leading approximation:
\begin{equation}
 D_r^{-1}(p^2)=\mu^2_r+p^2-g^2 L_r(p^2),
\end{equation}
where
$
 L_r(p^2)=L_0(p^2)-L_0(0)-p^2L'_0(0).
$

At $p^2\rightarrow \infty$,
\begin{equation}
D_r^{-1}(p^2)= (1-\frac{g_r^2}{96\pi^2m_r^2})p^2+O(\log\frac{p^2}{m_r^2}).
\label{D_MF_as}
\end{equation}
As can be seen from this equation, the asymptotic behavior of the
phion propagator
 in the deep Euclidean region is self-consistent
in the region of  weak coupling
which in this case is determined by the condition $g^2<g^2_c=96\pi^2m_r^2$.
When $g^2>g^2_c$, the asymptotics of the inverse propagator becomes negative,
which leads to the appearance of a Landau singularity for the chion propagator
in the Euclidean region, which in turn means a
violation of basic physical principles.
As was mentioned in the Introduction, the existence of
similar restrictions on the value of the coupling constant in the model
has been noted by many authors. Some authors (see, e.g., \cite{Rosenfelder},
\cite{Barro})
 believe self-evident that the presence of this kind of critical constants
reflects the metastability of the model. It seems to us,
however, that
   the presence of such a singularity
  means firstly  the non-applicability of the calculation method
in the strong coupling region and the need for more meaningful
non-perturbative approximations.

\section{Two-particle approximation}

 The system of  SDEs, generated by functional-differential
equation  (\ref{etaeq}), is an infinite set
of integral  equations for $n$-particle phion functions 
$Z_n\equiv \delta^n
Z/\delta\eta^n|_{\eta=0}$. 
 The first three SDEs are 
equations (\ref{D1}), (\ref{D2}) and (\ref{D3}).
  The $n$th SDE
is the $(n-1)$th derivative of SDE  (\ref{etaeq})
with the source
being switched off and includes a set of functions from
one-particle function  (phion propagator)  to the $(n+1)$-particle 
phion function.
In order  to obtain a sequence of closed systems of
equations, we proceed as follows. 
We call  "the $n$-particle approximation of the system of SDEs"
  the system of  $n$ SDEs, in which the first $n-1$
 equations are exact and the $n$th
SDE is truncated by omitting the $(n+1)$-particle function.
 It is evident that
the sequence of such approximations goes to the 
exact set of SDEs at $n\rightarrow\infty$.
The one-particle approximation is simply equation  (\ref{D1})
 without $Z_2$. This approximation
has a trivial solution which is a free propagator. 
The two-particle approximation  is a system of equation
  (\ref{D1})  and equation  (\ref{D2})  without $Z_3$:
\begin{equation}
 (m^2-\partial^2_x)Z_2\left(\begin{array}{cc}x&y\\x'&y'\end{array}\right)-
g^2\int dx_1\,D_c(x-x_1)\, 
Z_2\left(\begin{array}{cc}x_1&x_1\\x'&y'\end{array}\right)\,\Delta(x-y)
=\delta(x-y')\Delta(x'-y).
\label{D2PA}
\end{equation}
This  system of equations will be
the  object of the present investigation.

An alternative method to obtain this system
is a modification of the mean-field expansion of section 2.
 In this
modified mean-field expansion,  equations (\ref{D1}) and (\ref{D2PA})
are the basic equations of the expansion. A construction of such
modified mean-field expansion was performed for the model $\lambda(\phi^*\phi)^2$
in work \cite{Rochev11} and easily can be extended  to the considered
model.

In framework of the two-particle approximation. 
  $Z_2$  can be expressed as a functional of $\Delta$.
In fact,  equation (\ref{D2PA}) can be written as
\begin{equation}
 \int dx_1dy_1\,K\left(\begin{array}{cc}x&y\\x_1&y_1\end{array}\right)\,
Z_2\left(\begin{array}{cc}x_1&y_1\\x'&y'\end{array}\right)=
\Delta_c(x-y')\Delta(x'-y),
\end{equation}
where $K = K_{ab}$ at $\Delta_a=\Delta_c\equiv (m^2-\partial^2)^{-1}$
and $\Delta_b=\Delta$ (see 
(\ref{kernel})). In correspondence with equation (\ref{inverse_kernel}),
we obtain
\begin{eqnarray}
 Z_2\left(
\begin{array}{cc} x&y\\x'&y'\end{array}
\right) =
\Delta_c(x-y')\Delta(x'-y)+
\nonumber \\
+\int dx_1dx_2\,\Delta_c(x-x_1)\Delta(x'-x_2)\, 
f(x_1-x_2)\Delta(x_1-y)\Delta_c(x_2-y'),
\end{eqnarray}
where $f$ is given by equation  (\ref{feq}) and
\begin{equation}
 L(x-y)=\Delta_c(x-y)\Delta(y-x).
\end{equation}

Equations  (\ref{propchi})
and (\ref{D1}) with the taking into account above formulae
 and equation
(\ref{inv_k_prop}) give us
the system of equations for the chion and phion propagators.
In the momentum space, this system has the form
\begin{equation}
\cases{D^{-1}(p^2)=\mu^2+p^2-g^2L(p^2)
\cr
 \Delta^{-1}(p^2)=m^2+p^2-g^2K(p^2),}
\label{2PA}
\end{equation}
where
\begin{equation}
 L(p^2)=\int \frac{d^4q}{(2\pi)^4}\,\Delta_c(p+q)\Delta(q), \;\;
K(p^2)=\int \frac{d^4q}{(2\pi)^4}\Delta_c(p-q)D(q).
\end{equation}

The renormalization of equations of the two-particle 
approximation  is similar to the renormalization
of equations of the mean-field expansion with the imposition of 
normalization  conditions
 (\ref{normDelta}) and (\ref{normD}) on the renormalized
propagators. The system of renormalized equations for the chion
and phion propagators is:
\begin{equation}
\cases{D^{-1}_r(p^2)=\mu^2_r+p^2-g^2L_r(p^2)
\cr
 \Delta^{-1}_r(p^2)=m^2_r+p^2-g^2K_r(p^2),}
\label{2PAr}
\end{equation}
where
\begin{equation}
 L_r=L(p^2)-L(0)-p^2L'(0), \;
K_r=K(p^2)-K(0)-p^2K'(0).
\end{equation}

The system of equations (\ref{2PAr}) is a system of
non-linear integral equations, which is quite
difficult for the analytical study.
In the study of the asymptotic behavior 
 in the deep Euclidean region, one can make the approximation,
which
greatly simplifies  calculations, namely to replace
$\Delta_c$ by  the asymptotics  at $p^2\rightarrow\infty$, i.e.
by the massless propagator  $1/p^2$.
This massless integration approximation
  is quite usual in investigations in the deep Euclidean region 
\cite{Weinberg}.

Using the well-known
formula
\begin{equation}
 \int  \frac{d^4q}{(2\pi)^4}\,
\,\frac{ \Phi(q^2)}{(p-q)^2}=\frac{1}{16\pi^2}\bigg[\frac{1}{p^2}
\int_0^{p^2} \Phi(q^2)\, q^2\, dq^2+\int_{p^2}^\infty \Phi(q^2)\,dq^2\bigg]
\end{equation}
 gives us the system of  equations for 
propagators in the massless integration
approximation: 
\begin{equation}
 \cases{\Delta^{-1}(p^2)=m^2+(1-\frac{g^2}{32\pi^2\mu^2})p^2
+\frac{g^2}{16\pi^2}\int^{p^2}_0 dq^2 D(q^2)(1-\frac{q^2}{p^2})
\cr
D^{-1}(p^2)=\mu^2+(1-\frac{g^2}{32\pi^2 m^2})p^2
+\frac{g^2}{16\pi^2}\int^{p^2}_0 dq^2 \Delta (q^2)(1-\frac{q^2}{p^2}).}
\label{2PA_MLI}
\end{equation}
Here and  below we omit the index $r$, bearing in mind that
all quantities are renormalized.\\
Note that the system of equations (\ref{2PA_MLI}) is symmetric
with respect to the change
\begin{equation}
\Delta\longleftrightarrow D, \; m^2\longleftrightarrow\mu^2.
\label{sym}
\end{equation}
With taking into account 
 this symmetry, we introduce  the dimensionless variables: 
$$
u=\frac{\Delta^{-1}}{m^2},\;  v=\frac{D^{-1}}{\mu^2},\;
t=\frac{p^2}{\mu m},\; t'=\frac{q^2}{\mu m},\;
\lambda=\frac{g^2}{32\pi^2\mu m}.
$$
(Here $m\equiv\sqrt{m^2},\; \mu\equiv\sqrt{\mu^2}.$)

In terms of these variables, the system of equations (\ref{2PA_MLI})
takes the form
\begin{equation}
 \cases{u(t)=(\frac{\mu}{m}-\lambda)t+1+2\lambda\int_0^t \frac{dt'}{v(t')}(1-\frac{t'}{t})
\cr v(t)=(\frac{m}{\mu}-\lambda)t+1+2\lambda\int_0^t \frac{dt'}{u(t')}(1-\frac{t'}{t}).}
\label{system}
\end{equation}
The normalization conditions (\ref{normDelta}) and (\ref{normD})
for the dimensionless functions $u$ and $v$ have the form
\begin{equation}
 u(0)= v(0)=1; \; \; \dot{u}(0)=\frac{\mu}{m}, \; \dot{v}(0)=\frac{m}{\mu}.
\label{ic}
\end{equation}
From the system of integral equations (\ref{system}) one can easily
 go to the system of nonlinear differential equations for the functions $u$ and $v$:
\begin{equation}
 \cases{ \frac{d^2}{dt^2}(tu)=2(\frac{\mu}{m}-\lambda)+\frac{2\lambda}{v}
\cr \frac{d^2}{dt^2}(tv)=2(\frac{m}{\mu}-\lambda)+\frac{2\lambda}{u}. }
\label{diff_sys}
\end{equation}
The boundary conditions for this system are the normalization condition (\ref{ic}). 

If $v$ (or $u$)  increases in the absolute value at  $t\rightarrow\infty$, 
then for $\lambda\neq m/\mu, \; \lambda\neq \mu/m$, we obtain
from the system (\ref{diff_sys})
the  asymptotic behavior in the deep Euclidean region:
\begin{equation}
\cases{u = (\frac{\mu}{m}-\lambda)t +O(\log t ) \cr
v = (\frac{m}{\mu}-\lambda)t +O(\log t).}
\label{weak}
\end{equation}

This asymptotic behavior is self-consistent in
the weak-coupling region at \\
$\lambda<\mbox{ min}\{\frac{\mu}{m},\, 
\frac{m}{\mu} \}$ ( $\frac{g^2}{32\pi^2}<\mbox{ min}\{\mu^2,\, m^2 \})$
and corresponds to the asymptotically-free behavior of propagators:
\begin{equation}
 \Delta\sim 1/p^2,
\; D\sim 1/p^2.
\end{equation}

Outside the weak-coupling  region one should  distinguish three
cases: \\ $\mu^2=m^2$, $\mu^2>m^2$ and $\mu^2<m^2$.

At $\mu^2=m^2$, the system of equations (\ref{system}) takes the form:
\begin{equation}
 \cases{u=(1-\lambda)t+1+2\lambda\int_0^t \frac{dt'}{v(t')}(1-\frac{t'}{t})
\cr v=(1-\lambda)t+1+2\lambda\int_0^t \frac{dt'}{u(t')}(1-\frac{t'}{t}).}
\label{sys1}
\end{equation}
An iterative solution of this system gives us $u^{(n)} = v^{(n)}$  for
any finite number $n$ of iterations; hence, $u =v$ (provided
the convergence of the iteration procedure). Unfortunately, it is difficult
to prove the convergence of this nonlinear iteration scheme. Nevertheless, 
we will assume the equality of $u$ and $v$ 
 in the study of the case $\mu^2 = m^2$,
 leaving the possibility of an alternative
for a subsequent study.

At $u=v$, the system of equations (\ref{sys1}) is 
reduced to the single integral equation
\begin{equation}
 u=(1-\lambda)t+1+2\lambda\int_0^t \frac{dt'}{u(t')}(1-\frac{t'}{t}),
\label{eq_u}
\end{equation}
which,  in turn, can be  reduced  to the differential equation
\begin{equation}
 \frac{d^2}{dt^2}(tu)=2(1-\lambda)+\frac{2\lambda}{u}.
\label{eq_u_diff}
\end{equation}
 
In the strong-coupling region $\lambda>1$,  equation (\ref{eq_u_diff}) has
the positive exact solution
\begin{equation}
 u_{exact}=u_0=\frac{\lambda}{\lambda-1}.
\label{u_exact}
\end{equation}
Note that  $u_0$ is a solution  of the integral equation
\begin{equation}
 u_0=(1-\lambda)t+\frac{1}{1-\frac{1}{\lambda}}
+2\lambda\int_0^t \frac{dt'}{u_0}(1-\frac{t'}{t}),
\end{equation}
which differs   from equation (\ref{eq_u}) by the inhomogeneous term,
and this difference is small for large
$t$ in  comparision with  the leading part  $(1-\lambda)t$ 
of the inhomogeneous term.
This points  that $u_0$ is an asymptotic
solution of  equation (\ref{eq_u}) at $t\rightarrow\infty$.
In fact, performing the linearization:
\begin{equation}
 u=u_0+u_1, \;  \; \frac{1}{u}\simeq \frac{1}{u_0}-\frac{u_1}{u_0^2},
\label{linearization}
\end{equation}
we obtain for  $u_1$ the linear differential equation
\begin{equation}
 \frac{d^2}{dt^2}(tu_1)=-a^2u_1,
\label{eq_u1}
\end{equation}
where
\begin{equation}
 a=(\lambda-1)\sqrt{\frac{2}{\lambda}}.
\label{a}
\end{equation}
The solution of the linearized equation (\ref{eq_u1}) is
\begin{equation}
 u_1=\frac{1}{\sqrt{t}}\Big( A_1J_1(2a\sqrt{t})+A_2Y_1(2a\sqrt{t})\Big).
\label{u_1}
\end{equation}
Here $J_1$ и $Y_1$ are the cylindrical functions.\footnote{For
 exact solution of the linearized integral equation
see  Appendix.} Therefore, $u_1=o(u_0)$ at large $t$,
and constant $u_0$ is really the  asymptotic solution
in the strong-coupling region. 

At the critical value  $\lambda = 1$, the differential
equation (\ref{eq_u_diff}) has the exact solution
\begin{equation}
u_{exact}= \bar{u}=\sqrt{\frac{8t}{3}}.
\label{u_exact1}
\end{equation}
 $\bar{u}$ is the solution of  
integral equation
\begin{equation}
 \bar{u}=2\int_0^t \frac{dt'}{\bar{u}}(1-\frac{t'}{t}),
\end{equation}
which differs from equation (\ref{eq_u}) at $\lambda=1$ by
the inhomogeneous term. It is an indication that $\bar{u}$ 
is the asymptotic solution of equation (\ref{eq_u}) in
the critical point $\lambda=1$,
since this inhomogeneous term is  $o(\bar{u})$ at 
 $t\rightarrow\infty$. 
Indeed, 
performing  the linearization
\begin{equation}
 u=\bar{u}+u_1,\; \frac{1}{u}\simeq \frac{1}{\bar{u}}-\frac{u_1}{\bar{u}^2},
\end{equation}
we obtain the linearized equation for  $u_1$: 
\begin{equation}
 \frac{d^2}{dt^2}(tu_1)=-\frac{3}{4t}u_1,
\end{equation}
which is  an Euler equation, and the real solution of this equation is
\begin{equation}
 u_1=\frac{1}{\sqrt{t}}\Big(A_1\sin \frac{\log t}{\sqrt{2}}+
A_2\cos \frac{\log t}{\sqrt{2}}\Big).
\end{equation}
Therefore,  $u_1 =o(\bar{u})$
at $t\rightarrow\infty$.

Thus, the asymptotic behavior of  propagators at the critical point
$\lambda=1$
has the form
\begin{equation}
 \Delta\sim 1/\sqrt{p^2},
\; D\sim 1/\sqrt{p^2}.
\end{equation}

For unequal masses, 
we will consider  taking into account  the symmetry  (\ref{sym})
 only the case
$\mu^2> m^2$. In this case, there are two critical
couplings: $\lambda_{c1} = m/\mu$ and $\lambda_{c2} = \mu/m$, which
determine the weak-coupling region ($\lambda<m/\mu$), the intermediate
 region ($m/\mu <\lambda <\mu/m$) and the strong-coupling region ($\lambda> \mu/m $).

At $\lambda = \lambda_{c1} = m/\mu$, as in the 
weak-coupling region, there is a self-consistent solution with the asymptotic
behavior for large $t$ of the form
\begin{equation}
 u\simeq \frac{\mu^2-m^2}{\mu m}t, \;v\simeq \frac{2m^2}{\mu^2-m^2}\log t.
\end{equation}

In the strong-coupling region $\lambda> \lambda_{c2}$,
the system of differential equations (\ref{diff_sys}) has the
positive exact solution
\begin{equation}
  u_{exact}=u_0=\frac{\lambda\mu}{\lambda\mu-m},
\;\; v_{exact}=v_0=\frac{\lambda m}{\lambda m-\mu}. 
\end{equation}
$\{u_0, \,v_0\}$ is the solution 
of the system of integral equations
\begin{equation}
 \cases{u_0=(\frac{\mu}{m}-\lambda)t+\frac{1}{1-\frac{m}{\lambda\mu}}
+2\lambda\int_0^t \frac{dt'}{v_0}(1-\frac{t'}{t})
\cr v_0=(\frac{m}{\mu}-\lambda)t+\frac{1}{1-\frac{\mu}{\lambda m}}
+2\lambda\int_0^t \frac{dt'}{u_0}(1-\frac{t'}{t})}
\end{equation}
As in the above case of equal masses, this system
  differs from the system (\ref{system}) by inhomogeneous terms,
and this difference  is $O(1)$ at large $t$, 
while the inhomogeneous term is  $O(t)$. This
fact indicates that $\{u_0, \, v_0 \}$ is an asymptotic
solution at $t\rightarrow \infty$.

The linearization of the system of differential
equations (\ref{diff_sys}), completely analogous to the linearization of equation
(\ref{eq_u_diff}), leads to linear equations,
the solution of which, as well as for the case of $\mu^2 = m^2$, is expressed through
cylindrical functions and has the order of $o(1)$ for large $t$, which confirms
the conclusion that $\{u_0, \, v_0 \}$ is the leading asymptotic solution.

In the  intermediate region $m /\mu <\lambda <\mu /m$,
there are no
self-consistent solutions.

In fact, if either $u$,  or  $ v $ increase
in absolute value, then
the system of differential equations (\ref{diff_sys}) dictates
 asymptotic behavior  (\ref{weak}), 
i.e. the function $v$ is negative  at large
$t$ and, therefore,
 the chion propagator has a
Landau-type singularity.

If the function $v$ is positive
and  bounded, i.e. $0<v\leq v_{max}$, then
  from the integral equation for $u$ (see (\ref{system})), we obtain
$u\geq (\mu/m- \lambda + \lambda /v_{max})t +1$, i.e. the
function $u$ increases,
and this increase again
leads to the infinite increase (in the absolute value) of $v$
in contradiction with  the
assumption.

A similar analysis shows that at
$\lambda =\lambda_{c2} =\mu/m$ a self-consistent solution
  also does  no exist.

Thus, in the  case of unequal masses the critical point becomes a segment of
 critical values $\lambda_{c1}<\lambda \leq\lambda_{c2}$, in which
self-consistent solutions are not available.

\section{Three-particle approximation}

The two-particle approximation 
as well as the leading term of the  mean-field expansion
   has
an incomplete crossing structure of  
two-particle function. For  the two-particle approximation
the violation is even more significant  because it affects
 the disconnected
part.
However, in the same way as for  
the mean-field expansion, this problem is solved by considering the next
approximation. The next
three-particle approximation  is described by a system of three equations,
the first two of which are  equations (\ref{D1}) and (\ref{D2}), and the
third is  equation (\ref{D3}) without $Z_4$.

Just as  in the two-particle approximation, $Z_2$ 
can be expressed as a functional of $\Delta$,
in  the three-particle approximation 
 $Z_3$ can be expressed as a functional of $Z_2$ and $\Delta$:
\begin{equation}
Z_3\left(\begin{array}{ccc}x&y\\x'&y'\\x''&y''\end{array}\right)=
\int dx_1dy_1\,
K^{-1}\left(\begin{array}{cc}x&y\\x_1&y_1\end{array}\right)
Z_3^0\left(\begin{array}{ccc}x_1&y_1\\x'&y'\\x''&y''\end{array}\right),
\end{equation}
where 
 \begin{eqnarray}
Z_3^0\left(\begin{array}{ccc}x&y\\x'&y'\\x''&y''\end{array}\right)\equiv
-\bigg\{\Delta_c(x-y'')Z_2\left(\begin{array}{cc}x'&y\\x''&y\end{array}\right)+
\nonumber \\
+g^2\int dx_1dx_2\,
\Delta_c(x-x_1)\,D_c(x_1-x_2) 
Z_2\left(\begin{array}{cc}x_2&x_2\\x'&y'\end{array}\right)
\,Z_2\left(\begin{array}{cc}x''&y''\\x_1&y\end{array}\right)\bigg\}-
\nonumber \\
-\bigg\{x'\leftrightarrow x'', y'\leftrightarrow y''\bigg\},
\label{Z3_eq3PA}
\end{eqnarray}
and $K^{-1}$ is defined by equation  (\ref{inverse_kernel}) at
 $\Delta_a=\Delta_c, \;
\Delta_b=\Delta$.
Using this expression and the properties of the inverse
kernel $ K^{-1}$, we obtain an equation for $ Z_2 $ in 
the three-particle approximation, which can be written
as
\begin{equation}
 Z_2\left(\begin{array}{cc}x&y\\x'&y'\end{array}\right)=
 \bar{Z}_2\left(\begin{array}{cc}x&y\\x'&y'\end{array}\right)+
\int dx_1dy_1\,
\Delta_c(x-x_1)f(x_1-y_1)\Delta(x_1-y)
\bar{Z}_2\left(\begin{array}{cc}y_1&y_1\\x'&y'\end{array}\right),
\end{equation}
wher $\bar{Z}_2$ is the functional of  $Z_2$ and $\Delta$:
\begin{eqnarray}
\bar{Z}_2\left(\begin{array}{cc}x&y\\x'&y'\end{array}\right)
\equiv 
\Delta_c(x-y')\Delta(x'-y)+
\nonumber \\
+\int dx_1dx_2\,\Delta_c(x-x_1)f(x_1-x_2)\bigg\{
\Delta_c(x_2-y')Z_2\left(\begin{array}{cc}x_1&y\\x'&x_2\end{array}\right)
+\Delta_c(x_2-y)Z_2\left(\begin{array}{cc}x'&y'\\x_1&x_2\end{array}\right)+
\nonumber \\
+g^2\int dy_1dy_2\Delta_c(x_2-y_2)D_c(y_2-y_1)\bigg[
Z_2\left(\begin{array}{cc}y_1&y_1\\x_1&y\end{array}\right)
\,Z_2\left(\begin{array}{cc}x'&y'\\y_2&x_2\end{array}\right)+
\nonumber \\
+
Z_2\left(\begin{array}{cc}y_1&y_1\\x'&y'\end{array}\right)
\,Z_2\left(\begin{array}{cc}x_1&y\\y_2&x_2\end{array}\right)\bigg]\bigg\}
\end{eqnarray}
If we approximate in  $\bar{Z}_2 [Z_2, \Delta]$ the dependence on $Z_2$
by the
two-particle approximation, i.e. we consider the iterative solution, in which
$\bar{Z}_2 [Z_2, \Delta] \approx\bar{Z}_2 [Z_2^{2PA}, \Delta]$, then we can
show, using  equation  (\ref{D1}), that in this
approximation
 the correct
cross-symmetric structure is restored
for the disconnected part and also for
 the connected part of the two-particle
function.

\section{Conclusions}

The main result of the work  is the finding of 
 the  change of  asymptotic behavior
in the scalar Yukawa model 
in the framework of 
two-particle approximation.
The system of SDEs in the two-particle approximation
has   self-consistent 
solutions  in the weak-coupling
region (where a dominance of the perturbation theory is obvious)
 and in the strong-coupling region.
The field 
propagators
in the strong-coupling region asymptotically approach  constants. It 
is not something unexpected, if we look at the results of
 studying the strong-coupling region in  models of quantum field theory.
In particular, the well-known result in this direction is the conception
of the ultra-local approximation (also known as the ``static
ultra-local approximation''), considered in the paper
Caianiello and Scarpetta \cite{Caianiello}.
This exactly soluble approximation is 
based on removing of kinetic terms $\partial^2$ in the Lagrangian.
As a result,  all the Green functions are combinations of delta functions
in the coordinate space
 that are constants in the momentum space.
Of course, this approximation is very  difficult for a physical 
interpretation.
Nevertheless, it can be considered as a starting point for
 an expansion in inverse powers of the coupling constant, i.e.
as a leading approximation of the strong-coupling expansion 
\footnote{The ultra-local approximation and the 
strong-coupling expansion based on it are discussed in detail
 in the book of Rivers \cite{Rivers}. For 
later works, see   \cite{Klauder}, \cite{Svaiter}. For the ultra-local
approximation in the  bilocal source  formalism,
  see  \cite{Rochev93}.}.

In this connection, it is noteworthy  that the obtained  solutions
of the two-particle approximationin in the strong-coupling region for large
Euclidean momenta tend to constants. Such behavior  asymptotically
corresponds to the  ultra-local approximation.
On the other hand, our solutions are
free from the interpretation problems, since for the small
momenta they  have the quite traditional pole behavior.
This indicates that this 
approximation seems to be adequately described
as the weak-coupling as the strong-coupling too.

In the case  of equal masses, of fields a self-consistent  solution of 
 SDEs  in two-particle approximation exists for
 any value of the coupling, including the critical value.
At $g^2=g^2_c$, the asymptotics of propagators are 
 $1/p$, i.e. the asymptotic behavior is a medium among
the free behavior     $1/p^2$ at $g^2<g^2_c$ 
and the constant-type behavior in strong coupling region  $g^2>g^2_c$.
A sharp change of  asymptotic behavior in the vicinity of the critical
value is a behavior that is characteristic for a phase transition.
This phase transition is not associated with a symmetry breaking, and in 
this sense is similar to the phase transition of "gas--liquid".
The weak-coupling region can be roughly classified as the gaseous phase and
the strong-coupling region, where  a kind of
  localization of correlators exists -- to the liquid.
This analogy, of course, is a quite shallow.
A type and characteristics of this phase transition
  can be defined as the result of a detailed  study using methods
of the theory of critical behavior.

In the case  of unequal masses,  two critical
values of
   coupling exist.
In the interval between them, it is not possible to construct 
a solution 
with  self-consistent ultraviolet behavior.
Reasons for the existence of such an interval is currently unclear.
The existence of such intermediate values of coupling can 
somehow reflect the metastability of the model.
On the other  hand,  it may be  
 an artifact of  the two-particle
approximation.
In any case, a detailed study of 
the three-particle approximation of  Section 4
 will help to shed a light on this issue.

\section*{Appendix}

A linearization of  nonlinear differential 
equations in the strong-coupling region
has a sense, in general, only for large  $t$.
However, in  the particular case of equation
(\ref{eq_u_diff}) at $\lambda>1$, this linearization
available throughout the range of the $t$, starting from zero,
i.e. a linearization  of the integral equation
(\ref{eq_u}) is  possible. We can
use this fact to estimate the values
of constants $A_1$ and $A_2$ in equation (\ref{u_1}). 
Performing the
linearization of  integral equation (\ref{eq_u})
by formula (\ref{linearization}), we obtain for $u_1$
the linear integral equation
\begin{equation}
 u_1=-\frac{1}{\lambda-1}
- a^2\int_{0}^{t} dt'u_1(t')(1-\frac{t'}{t}),
\label{u_1_integral}
\end{equation}
where  $a$ is defined by equation (\ref{a}).
Equation
(\ref{u_1}) tells us to deal with the Ansatz
 in the form
\begin{equation}
 u_1=\frac{A_1}{\sqrt{t}}\,J_1(2a\sqrt{t}),
\label{u_1_anzats}
\end{equation}
i.e. $A_2=0$. 
The value of constant $A_1$ is determined by the
direct substitution of the expression (\ref{u_1_anzats})
into  integral equation (\ref{u_1_integral}). Calculating
integrals with the known formulae (see, e.g. \cite {Bateman}),
we obtain
\begin{equation}
 A_1=-\frac{1}{(\lambda-1)a}.
\label{A_1}
\end{equation}
Thus, the solution of the linearized approximation is
\begin{equation}
 u=u_0+u_1=\frac{\lambda}{\lambda-1}-\frac{1}{(\lambda-1)a\sqrt{t}}\,J_1(2a\sqrt{t}).
\end{equation}

\end{document}